\documentclass[twocolumn,showpacs,preprintnumbers,amsmath,amssymb]{revtex4}

\usepackage[dvips]{graphics}
\usepackage{amsmath, amsthm, amssymb}
\usepackage{graphicx}
\usepackage{dcolumn}
\usepackage{bm}

\begin{document}

\title{Probing the Electronic Structure of Bilayer Graphene by Raman Scattering}

\author{L.~M.~Malard$^1$, J. Nilsson$^2$, D. C. Elias$^1$, J. C. Brant$^1$, F. Plentz$^1$, E. S. Alves$^1$,
A. H. Castro Neto$^2$ and M. A. Pimenta$^1$}
\address{$^1$Departamento de F\'{\i}sica, Universidade Federal de Minas
Gerais, 30123-970, Belo Horizonte, Brazil \\ $^2$Department of
Physics, Boston University, 590 Commonwealth Avenue, Boston, MA
02215, USA}

\date{\today}

\begin{abstract}

The electronic structure of bilayer graphene is investigated from a
resonant Raman study using different laser excitation energies. The
values of the parameters of the Slonczewski-Weiss-McClure model for
graphite are measured experimentally and some of them differ
significantly from those reported previously for graphite, specially
that associated with the difference of the effective mass of
electrons and holes. The splitting of the two TO phonon branches in
bilayer graphene is also obtained from the experimental data. Our
results have implications for bilayer graphene electronic devices.

\end{abstract}

\maketitle

After the development of the phenomenological
Slonczewski-Weiss-McClure (SWM) model for the electronic structure
of 3D graphite 50 years ago \cite{McClure1957, SW1958}, many
experimental works were devoted to evaluate the parameters of this
standard model \cite{Millie88, graphitereview1}. In particular, the
weak interaction between two layers in graphite, ascribed to van der
Waals forces, can be described by the interlayer SWM parameters,
that give, for example, the difference between the effective masses
of electrons and holes near the Dirac point (K point in the graphite
Brillouin zone) \cite{Millie88}. This work shows that, by performing
Raman scattering experiments in bilayer graphene with many different
laser excitation energies, we can probe its electronic structure by
obtaining experimental values for the SWM parameters. Our values for
some interlayer parameters differ significantly from values measured
previously for  graphite \cite{Millie88, graphitereview1}, showing
that open issues about graphite can now be experimentally revisited
using graphene samples \cite{geim07}. Furthermore, while the
unbiased bilayer graphene is a zero-gap semiconductor, a biased
bilayer is a tunable gap semiconductor by electric field effect
\cite{castro07, vandersypen07}. Hence, the development of bilayer
graphene-based bulk devices depends on the detailed understanding of
its electronic properties.

\begin{figure}
\includegraphics [scale=0.5]{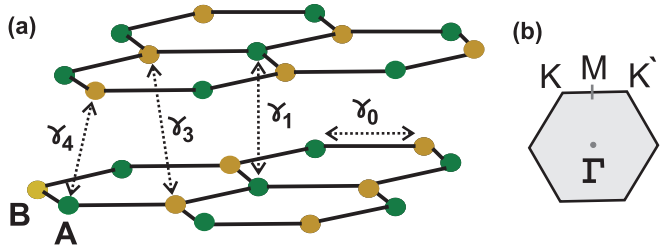}
\caption{\label{Fig1} (a) Atomic structure of bilayer graphene. The
A atoms of the two layers are over each other, whereas the B atoms
of the two layers are displaced with respect to each other. The SWM
constants $\gamma_0$, $\gamma_1$, $\gamma_3$ and $\gamma_4$ label
the corresponding pair of atoms associated with the hopping
processes. (b) First Brillouin zone of monolayer graphene, showing
the high symmetry points $\Gamma$, K, K$^\prime$ and M.}
\end{figure}

Fig.~\ref{Fig1} shows the atomic structure of a bilayer graphene, in
which we can distinguish the two nonequivalent atoms A and B in each
plane giving rise to a unit cell with four atoms. Since this unit
cell is the same for graphite in the Bernal stacking structure, we
can describe the electronic spectrum of bilayer graphene in terms of
the SWM model for graphite \cite{McClure1957, SW1958}, by
determining the parameters $\gamma_0$, $\gamma_1$, $\gamma_3$ and
$\gamma_4$, that are associated with overlap and transfer integrals
calculated for nearest neighbors atoms. The pair of atoms associated
with these parameters are indicated in the atomic structure of a
bilayer graphene shown in Fig.~\ref{Fig1}(a). These parameters, that
are fundamental for the electronic processes in the system, are only
roughly known to this date.

The graphene samples used is this experiment were obtained by a
micro-mechanical cleavage of graphite on the surface of a Si sample
with a 300 nm layer of SiO$_2$ on the top \cite{geim07}. The bilayer
flakes were identified by the slightly color change from monolayer
graphene in an optical microscope, followed by a Raman spectroscopy
characterization using the procedure described by Ferrari {\it et
al.} \cite{ferrari}. For the Raman measurements, we used a Dilor XY
triple monochromator in the back scattering configuration. The spot
size of the laser was $\sim$1 $\mu$m using a 100 $\times$ objective
and the laser power was kept at 1.2 mW in order to avoid sample
heating. Raman spectra were obtained for 11 different laser lines of
Ar-Kr and dye lasers in the range 1.91--2.71 eV.

\begin{figure}
\includegraphics [scale=0.7]{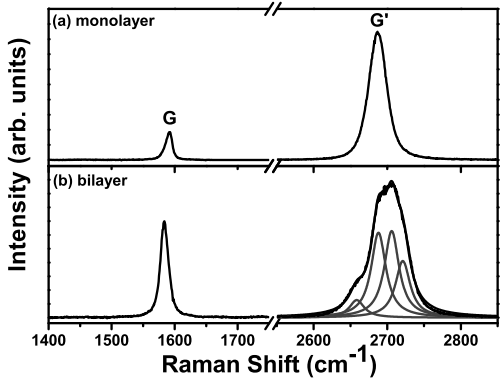}
\caption{\label{Fig2} (a) Raman spectrum of the monolayer graphene
and (b) Raman spectrum of the bilayer graphene performed with the
$2.41$ eV laser line. The G$^{\prime}$ band of bilayer graphene is
fit with four different Lorentzians with FWHM of $\sim$ $24$
cm$^{-1}$.}
\end{figure}

Recently, Ferrari {\it et al.} \cite{ferrari} showed that Raman
spectroscopy can be used to identify the number of layers in a
graphene sample and, in particular, to clearly distinguish a
monolayer from a bilayer graphene sample. Figure~\ref{Fig2} shows
the Raman spectra of the monolayer [Fig.~\ref{Fig2}(a)] and bilayer
[Fig.~\ref{Fig2}(b)] graphene samples, where the most prominent
features are the G and G$^\prime$ Raman bands \cite{pimenta07}. The
G$^\prime$ band of the monolayer graphene is nicely fitted by just
one Lorentzian, whereas four Lorentzian peaks are necessary to fit
the G$^\prime$ band of the bilayer graphene, in agreement with the
previous Raman studies of graphene systems \cite{ferrari, gupta06,
graf07}.

\begin{figure}
\includegraphics [scale=0.7]{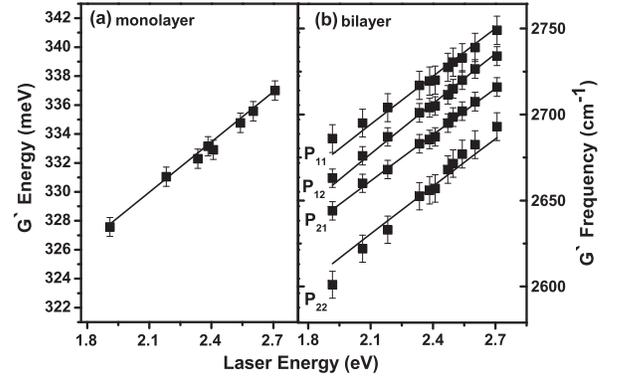}
\caption{\label{Fig3} (a) Laser energy dependence of the
G$^\prime$-band energy for a monolayer graphene. (b) Laser energy
dependence of the positions of the four peaks that comprise the
G$^\prime$-band of bilayer graphene. These four Raman peaks
originate from the P$_{11}$, P$_{12}$, P$_{21}$ and P$_{22}$ DR
processes illustrated in  Fig. \ref{Fig4}. The black squares are the
experimental data and the full lines are the fitted curves (see
discussion in the text).}
\end{figure}

The Raman spectra of both the monolayer and bilayer graphene have
been measured with many different laser energies in the visible
range. Figure~\ref{Fig3} shows the laser energy dependence of the
G$^{\prime}$-band frequency for the monolayer sample
[Fig.~\ref{Fig3}(a)] and for each one of the four peaks that
comprise the G$^{\prime}$ band for bilayer graphene
[Fig.~\ref{Fig3}(b)].

The origin of the G$^\prime$ band in all graphitic materials is due
to an inter-valley double-resonance (DR) Raman process involving
electronic states near two nonequivalent K points (K and K$^\prime$)
in the 1$^{st}$ Brillouin zone of graphene [see Fig. \ref{Fig1}(b)],
and two phonons of the iTO branch \cite{thomsen, saitoDR,
pimenta07}. As a result of the angular dependence of the
electron-phonon matrix elements \cite{georgii-thesis} and the
existence of interference effects in the Raman cross-section
\cite{maultzschPRB2004}, the main contribution for the
G$^\prime$-band comes from the particular DR process that occurs
along the $\Gamma$-K-M-K$^\prime$-$\Gamma$ direction, in which the
wavevectors $k$ and $k^\prime$ of the two intermediate electronic
states in the conduction band (measured from the K and K$^\prime$
points, respectively) are along the K$\Gamma$ and K$^\prime \Gamma$
directions, respectively \cite{ferrari}. Therefore, the wavevector
$q$ of the phonons involved in this specific process is along the KM
direction and is related to the electron wavevectors by the
condition $q=k+k^\prime$ \cite{pimenta07}.

\begin{figure}
\includegraphics [scale=0.35]{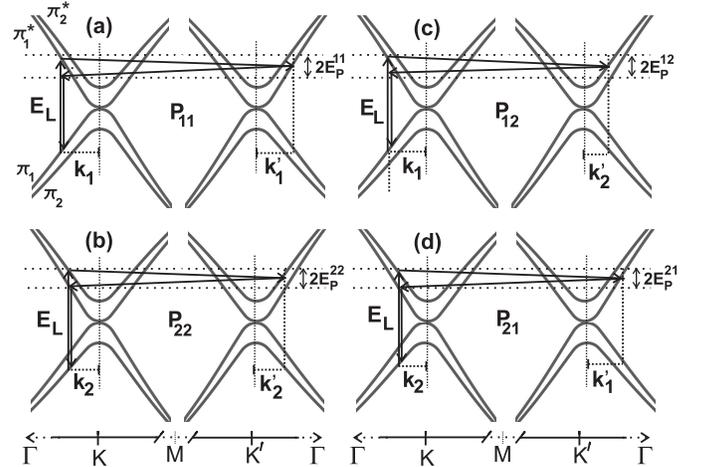}
\caption{\label{Fig4} Schematic view of the electron dispersion of
bilayer graphene near the K and K$^\prime$ points showing both
$\pi_{1}$ and $\pi_{2}$ bands. The four DR processes are indicated:
(a) process P$_{11}$, (b) process P$_{22}$, (c) process P$_{12}$,
and (d) process  P$_{21}$. The wavevectors of the electrons ($k_1$,
$k_2$, $k_1^\prime$ and $k_2^\prime$) involved in each of these four
DR processes are also indicated.}
\end{figure}

In the case of the bilayer graphene, where the two graphene layers
are stacked in a Bernal configuration \cite{pimenta07}, the
$\pi$-electrons dispersion in the valence and in the conduction band
splits into two parabolic branches near the K point \cite{mccann},
as shown in Fig.~\ref{Fig4}. In this figure, the upper and lower
branches of the valence band are labeled as $\pi_{1}$ and $\pi_{2}$,
respectively. The lower and upper branches of the conduction band
are called $\pi_{1}^*$ and $\pi_{2}^*$, respectively. Along the high
symmetry $\Gamma$-K-M-K$^\prime$-$\Gamma$ direction, these branches
belong to different irreducible representations of the P6$_3$/mmc
space group and, therefore, only the
$\pi_{1}$$\rightleftarrows$$\pi_{1}^*$ and
$\pi_{2}$$\rightleftarrows$$\pi_{2}^*$ optical transitions between
the valence and conduction bands are allowed. Therefore, there are
four possible intervalley DR processes involving electrons along the
$\Gamma$-K-M-K'-$\Gamma$ direction that lead to the observation of
the four peaks of the G$^\prime$-band of bilayer graphene
\cite{ferrari}.

The four DR processes are represented in Fig.~\ref{Fig4}. In process
P$_{11}$ [Fig.~\ref{Fig4}(a)], an electron with wavevector $k_1$ is
resonantly excited from the valence band $\pi_{1}$ to the conduction
band $\pi^*_{1}$ by absorbing a photon with energy $E_L$. This
electron is then resonantly scattered to a state with wavevector
$k^\prime_1$ by emitting a phonon with momentum q$_{11}$ and energy
$E^{11}_{p}$. Finally, the electron is scattered back to state $k_1$
by emitting a second phonon, and it recombines with a hole producing
a scattered photon with an energy $E_S=E_L-2E^{11}_p$. The phonon
wavevector q$_{11}$, measured from the K point and along the KM
direction, is given by q$_{11}$ = $k_1$ + $k^{\prime}_1$.

Fig.~\ref{Fig4}(b) shows the DR process P$_{22}$, which involves an
electron that is optically excited between the $\pi_2$ and $\pi^*_2$
branches. The energy of the associated phonon is $E^{22}_{p}$ and
its wavevector is given by q$_{22}$ = $k_2$ + $k^{\prime}_2$.
Figures~\ref{Fig4}(c) and \ref{Fig4}(d) show processes P$_{12}$ and
P$_{21}$, that involve electrons with wavevectors  $k_1$,
$k^\prime_2$ and $k_2$, $k^\prime_1$, respectively, which belong to
different electron branches. The wavevectors of the phonons
associated with processes P$_{12}$ and P$_{21}$ are given by
q$_{12}$ = $k_1$ + $k^{\prime}_2$ and q$_{21}$ = $k_2$ +
$k^{\prime}_1$, respectively.

The two iTO phonon branches of a bilayer graphene along the
$\Gamma$KM line belong to the irreducible representations T$_{1}$
and T$_{3}$. The scattering of an electron in the conduction band
between states around  K and K$^\prime$ has to satisfy the
electron-phonon selection rule, so that the allowed transitions are
$\pi_{1}^*$$\rightleftarrows$$\pi_{1}^*$ or
$\pi_{2}^*$$\rightleftarrows$$\pi_{2}^*$ for the T$_{1}$ phonon and
$\pi_{2}^*$$\rightleftarrows$$\pi_{1}^*$ for the T$_{3}$ phonon.
Since processes P$_{11}$ and P$_{22}$ involves electrons states
around K and K$^\prime$ which belong to the same electronic branch,
the associated phonons belong to the T$_{1}$ phonon branch. On the
other hand, phonons involved in processes P$_{12}$ and P$_{21}$
belong to the T$_{3}$ branch.

From Fig.~\ref{Fig4} one can see that the phonon associated with the
P$_{11}$ process has the largest wavevector (q$_{11}$). Since the
energy of the iTO phonon increases with increasing $q$, the highest
frequency component of the G$^\prime$-band of a bilayer graphene is
related to the P$_{11}$ process. On the other hand, P$_{22}$ process
is associated with the smallest phonon wavevector q$_{22}$, and
gives rise to the lowest frequency component of the G$^\prime$-band
of a bilayer graphene. The two intermediate peaks of the
G$^\prime$-band are associated with processes P$_{12}$ and P$_{21}$.

In order to analyze the experimental results shown in
Fig.~\ref{Fig3}(b), we must find a relation between the electronic
and the phonon dispersions of a bilayer graphene, according to the
DR Raman process. The electronic dispersion of the bilayer graphene
can be described in terms of the standard SWM model for graphite
(see e.g. Eq.~(2.1) in Ref.~\cite{McClure1957}) and using the full
tight-binding dispersion introduced originally by Wallace
\cite{Wallace47}. Along the $K-\Gamma$ direction this amounts to
replacing $\sigma$ by $\gamma_0 \bigl[ 2 \cos(2 \pi/3 - k a
\sqrt{3}/2 ) + 1 \bigr]$ in McClure's expressions. Here $k$ is
measured from the $K$ point and $a = 1.42 \,\text{\AA}$ is the in
plane nearest neighbor carbon distance. Since there is no
$k_z$-dependence we may set $\Gamma=1$ and  $\gamma_2 = \gamma_5 =
0$ ($\gamma_2$ and $\gamma_5$ correspond to 3$^{rd}$ layer
interactions in graphite). We have verified that the parameter
$\Delta $ does not make any noticeable difference in our results and
will be ignored in our analysis. With these simplifications, the
bands in the bilayer are parameterized by the four parameters
$\gamma_0$, $\gamma_1$, $\gamma_3$ and $\gamma_4$. A more detailed
account on this approach to the band structure calculation of the
graphene bilayer can be found in Ref.~\cite{Partoens2006}.

In fact, along the high symmetry  $K-\Gamma$ direction, the $4
\times 4$ matrix factorizes and the dispersion of the four bands are
given by:
\begin{subequations}
\begin{eqnarray}
  \label{eq:band_dispersion}
  E_{\pi_2} &=&  (-\gamma_1 - v_3 \sigma - \xi_{+} ) / 2 ,
  \\
  E_{\pi_1} &=&  ( \gamma_1 + v_3 \sigma - \xi_{-} ) / 2,
  \\
  E_{\pi_1^*} &=&  (-\gamma_1 - v_3 \sigma + \xi_{+} ) / 2,
  \\
  E_{\pi_2^*} &=&  ( \gamma_1 + v_3 \sigma + \xi_{-} ) / 2,
\end{eqnarray}
\end{subequations}
where
\begin{equation}
  \label{eq:Eplusminus}
  \xi_{\pm}
  = \sqrt{(\gamma_1 - v_3 \sigma )^2
    +4  (1 \pm v_4)^2 \sigma^2  },
\end{equation}
and $v_j \equiv \gamma_j / \gamma_0$.

In order to obtain the dependence of the phonon energy $E^{ij}_p$ on
the laser energy $E_L$, let us consider a generic process P$_{ij}$,
where $i$, $j$ = 1 or 2, which describes the four processes shown in
Fig.~\ref{Fig4}. In the first step of this process (electron-hole
creation), the incident photon is in resonance with the electronic
states in the valence and conduction bands at the $k_i$ point. Thus,
the laser energy $E_L$ can be written as:
\begin{equation}\label{eq2}
E_L=E_{\pi_i^*}(k_i)-E_{\pi_i}(k_i).
\end{equation}
Eq. \ref{eq2} allows us to determine the momentum $k_i$ of the
electron excited in the process.

The electron is then resonantly scattered from the vicinity of the K
point to the vicinity of the K$^\prime$ point by emitting a (iTO)
phonon with energy $E^{ij}_p$ given by:
\begin{equation}\label{eq3}
E^{ij}_p(k_i+k^{\prime}_j)= E_{\pi_i^*}(k_i) -
E_{\pi_j^*}(k^{\prime}_j)
\end{equation}
Assuming that we know the iTO phonon dispersion near the K point
$[A+B(k_i+k^{\prime}_j)]$ as well as the bands involved (Eqs.~1 and
\ref{eq:Eplusminus}), Eq. \ref{eq3} uniquely determine the momentum
of the scattered electron $k^{\prime}_j$. We then compute the phonon
energy $E^{ij}_p$, that is directly related to the Raman shift for
this specific P$_{ij}$ process, obtained with a given laser energy
$E_L$. Finally, we perform a least-squares fit to determine the
parameters $\gamma_0$, $\gamma_1$, $\gamma_3$ and $\gamma_4$ of the
model (Eqs.~1 and \ref{eq:Eplusminus}) that give the best fit of the
dispersion of the four G$^\prime$ peaks in bilayer graphene shown in
Fig.~\ref{Fig3}. The four solid lines in Fig.~\ref{Fig3}(b)
represent the best fitting of the experimental $E^{ij}_p$ versus
$E_L$ data.

We have also tried to fit, unsuccessfully, our experimental data for
the four DR processes taking only $\gamma_0$, $\gamma_1$ and
$\gamma_3$. Therefore, in order to get a good fitting, the parameter
$\gamma_4$ value, that is associated with the splitting of the two
G$^{\prime}$ intermediate peaks, has to be included. For the
dispersion of the iTO phonon branches near the K point, we could not
fit satisfactorily the dispersion data in Fig.~\ref{Fig3}(b)
considering the same phonon dispersion for the four P$_{ij}$
processes. The best fit was obtained when we considered different
dispersions for the two iTO phonon branches of bilayer graphene.
Table \ref{phonon} shows the parameters obtained for each phonon
branch, which exhibit a linear dispersion near the K point. Notice
that the difference between these two phonon branches corresponds to
a splitting of about 3 cm$^{-1}$, which is in close agreement to
that reported previously \cite{ferrari, graf07}.

\begin{table}[htbp]
\begin{footnotesize}
 \caption
  {Values obtained for the phonon dispersion of the two iTO phonon
  branches of bilayer graphene near the K point, where
 $E^{ii}_p$=A$_{ii}$+B$_{ii}$q corresponds to the P$_{11}$ and P$_{22}$
    processes
    and $E^{ij}_p$=A$_{ij}$+B$_{ij}$q to the P$_{12}$ and
    P$_{21}$ processes.}
 \label{phonon}
  \centering

  \begin{tabular}{|l|l|l|l|}
    \hline
     $A_{ii}  $ & $B_{ii}  $
    & $ A_{ij}  $ & $B_{ij}  $ \\
    \hline
    153.7 (\text{meV}) & 38.5 (\text{meV\,\AA})& 154.0 (\text{meV})& 38.8 (\text{meV\,\AA})\\
    1238 (\text{cm$^{-1}$}) & 310 (\text{cm$^{-1}$\,\AA})& 1241 (\text{cm$^{-1}$})& 313 (\text{cm$^{-1}$\,\AA})\\
    \hline
  \end{tabular}
 \end{footnotesize}
\end{table}

Table \ref{tbparameters} shows the $\gamma$ values obtained
experimentally. The parameter $\gamma_0$, associated with the
in-plane nearest-neighbor hopping energy, is ten times larger than
$\gamma_1$, which is associated with atoms from different layers
along the vertical direction (see Fig.~\ref{Fig1}). These values are
in good agreement with the previous angle resolved photoemission
spectroscopy (ARPES) measurements in bilayer graphene
\cite{ohta2007}. The resolution of our experiment allows, however,
the measurement of weaker hopping parameters ($\gamma_3$ and
$\gamma_4$) , that are beyond the current resolution of ARPES. The
value of $\gamma_1$ is about three times larger than $\gamma_3$ and
$\gamma_4$, both associated with the interlayer hopping not along
the vertical direction (see Fig.~\ref{Fig1}).

The corresponding parameters found experimentally for graphite are
also shown in Table \ref{tbparameters} \cite{Millie88,
graphitereview1}. The parameters $\gamma_0$ and $\gamma_1$ for
bilayer graphene are slightly smaller than those for graphite. This
difference is more accentuated for the parameter $\gamma_3$. Notice
that our value for $\gamma_3$ is in good agreement with that
reported theoretically by Tatar and Rabii \cite{tatar82}. On the
other hand, our value of $\gamma_4$ for bilayer graphene is
significantly higher than the value for graphite measured by Mendez
\emph{et al.} in a magnetoreflection experiment. \cite{mendez80}.
Moreover, this parameter is specially important since it is related
to the difference of electron and hole effective masses in the
valence and conduction bands \cite{Millie88}. This is particularly
important in the context of the biased bilayer graphene
\cite{castro07}. We must stress that our value for $\gamma_4$ is in
good agreement with the calculations performed by Tatar and Rabii
\cite{tatar82} and by Charlier \emph{et al.} \cite{charlier92}.

\begin{table} [htbp]
\begin{footnotesize}
\caption
  {Experimental SWM parameters (in eV)
    for the band structure of bilayer graphene. The parameters for graphite
are taken from Refs.~\onlinecite{graphitereview1, Millie88}.}
\label{tbparameters}
  \centering
  \begin{tabular}{|l|l|l|l|l|l|l|l|l|}
    \hline  &
    $\gamma_0$ & $\gamma_1$ & $\gamma_2$ & $\gamma_3$ &
    $\gamma_4$ & $\gamma_5$ & $\gamma_6 = \Delta$ \\
    \hline  Bilayer graphene &
    2.9 & 0.30 & n/a & 0.10 & 0.12 & n/a & n/a \\
    \hline Graphite &
    3.16 & 0.39 & -0.02 & 0.315 & 0.044 & 0.038 & 0.008 \\
    \hline
  \end{tabular}
  \end{footnotesize}
\end{table}

In summary, from the resonant Raman study of the G$^\prime$-band of
bilayer graphene using several laser excitation energies, we have
been able to probe the dispersion of electrons and phonons of this
material near the Dirac point. From the fitting of the experimental
data using the SWM model, we have obtained experimental values for
hopping parameters for bilayer graphene. The parameters $\gamma_3$
and $\gamma_4$ found here differ significantly from the values
previously measured for graphite \cite{Millie88, mendez80,
graphitereview1}, showing that open issues about graphite can now be
revisited using graphene samples. The slight difference between the
experimental and theoretical data in Fig. \ref{Fig3}(b), specially
for lower laser energies, might be ascribed to many-body effects in
bilayer graphene \cite{viola07}. Finally, in a future work we will
study DR processes involving holes as well as the biased bilayer
graphene.

This work was supported by Rede Nacional de Pesquisa em Nanotubos de
Carbono - MCT, Brasil and FAPEMIG. L.M.M., J.C.B and D.C.E.
acknowledge the support from the Brazilian Agency CNPq. We would
like to thank Millie Dresselhaus and Ado Jorio for critical reading
of the manuscript and for helpful discussions.

\end{document}